\documentclass[sigconf]{acmart} 
\AtBeginDocument{%
  }

\usepackage[utf8]{inputenc} 
\usepackage[T1]{fontenc}    
\usepackage{hyperref}       
\usepackage{url}            
\usepackage{booktabs}       
\usepackage{amsfonts}       
\usepackage{nicefrac}       
\usepackage{microtype}      

\usepackage[utf8]{inputenc} 
\usepackage[T1]{fontenc}    
\usepackage{hyperref}       
\usepackage{pifont}
\usepackage{url}            
\usepackage{booktabs}       
\usepackage{amsfonts}       
\usepackage{nicefrac}       
\usepackage{microtype}      
\usepackage{lipsum}
\usepackage{graphicx}       
\graphicspath{{media/}}     
\usepackage{mathtools}
\usepackage{amsmath,amsfonts} 

\usepackage{graphicx}
\usepackage{textcomp}
\usepackage{xcolor}
\usepackage{comment}
\usepackage[T1]{fontenc}
\usepackage[utf8]{inputenc}
\usepackage{verbatim}
\usepackage{soul} 
\usepackage[hyphenbreaks]{breakurl}
 \usepackage{float}
\usepackage{hyperref}
\usepackage[hyphenbreaks]{breakurl}
\usepackage{multirow}
\usepackage{adjustbox}

\usepackage{algorithm}
\usepackage{algorithmic}

\usepackage{enumerate}
\usepackage{colortbl}
\usepackage{booktabs}
\usepackage{array}
\usepackage[english]{babel}
\usepackage{amsthm}
\newtheorem{theorem}{Theorem}[section]
\newtheorem{lemma}[theorem]{Lemma}
\usepackage{setspace}

\usepackage{framed}
\usepackage{mdframed}

\theoremstyle{plain}

\usepackage{mdframed}
\definecolor{theoremcolor}{rgb}{0.97, 0.97, 0.97} 
\definecolor{examplecolor}{rgb}{1, 1, 1.0}
\mdfsetup{
    innertopmargin=5pt,
    innerbottommargin=5pt,
    leftmargin=3pt,
    rightmargin=3pt,
    backgroundcolor=theoremcolor,
    linewidth=0pt,
}
\usepackage{xfrac}
\usepackage{comment}

\newmdtheoremenv{prop}{Proposition}

\theoremstyle{remark}
\newtheorem{remark}[theorem]{Remark}

\usepackage{tikz}
\usetikzlibrary{shadows}

\usepackage{algorithm}
\usepackage{algorithmic}
\usepackage{threeparttable}
\usepackage[table,xcdraw]{xcolor}
\usepackage[normalem]{ulem}
\usepackage{wrapfig}
\usepackage{balance}
\useunder{\uline}{\ul}{}
\usepackage{xspace}

\usepackage[most]{tcolorbox}

\usepackage{etoolbox}
\AtBeginEnvironment{equation}{\setlength{\abovedisplayskip}{2pt}\setlength{\belowdisplayskip}{2pt}}
\AtBeginEnvironment{equation*}{\setlength{\abovedisplayskip}{2pt}\setlength{\belowdisplayskip}{2pt}}
\AtBeginEnvironment{align}{\setlength{\abovedisplayskip}{2pt}\setlength{\belowdisplayskip}{2pt}}
\AtBeginEnvironment{align*}{\setlength{\abovedisplayskip}{2pt}\setlength{\belowdisplayskip}{2pt}}

\copyrightyear{2026}
\acmYear{2026}
\setcopyright{cc}
\setcctype{by}
\acmConference[WSDM '26]{Proceedings of the Nineteenth ACM International Conference on Web Search and Data Mining}{February 22--26, 2026}{Boise, ID, USA}
\acmBooktitle{Proceedings of the Nineteenth ACM International Conference on Web Search and Data Mining (WSDM '26), February 22--26, 2026, Boise, ID, USA}
\acmPrice{}
\acmDOI{10.1145/3773966.3779367}
\acmISBN{979-8-4007-2292-9/2026/02}

\acmConference[WSDM '26]{Web Search and Data Mining}{22 Feb - 26 Feb 2026}{Boise, Idaho, USA}

\begin{document}

\title{Factorized Transport Alignment for Multimodal and Multiview E-commerce Representation Learning}

\author{Xiwen Chen}
\authornote{Work done at Etsy during summer internship.}
\email{xiwenc@g.clemson.edu}
\affiliation{%
  \institution{Clemson University}
  \city{Clemson}
  \state{SC}
  \country{USA}
}

\author{Yen-Chieh Lien}
\author{Susan Liu}
\author{María Castaños}
\email{{ylien,susanliu,mcastanos}@etsy.com} %
\affiliation{%
  \institution{Etsy, Inc.}
  \city{Brooklyn}
  \state{NY}
  \country{USA}
}

\author{Abolfazl Razi}
\email{arazi@clemson.edu}
\affiliation{%
  \institution{Clemson University}
  \city{Clemson}
  \state{SC}
  \country{USA}
}

\author{Xiaoting Zhao}
\author{Congzhe Su}
\email{{xzhao,csu}@etsy.com} %
\affiliation{%
  \institution{Etsy, Inc.}
  \city{Brooklyn}
  \state{NY}
  \country{USA}
}
\renewcommand{\shortauthors}{Xiwen Chen et al.}

\renewcommand{\shortauthors}{Xiwen Chen et al.}

\begin{abstract}
The rapid growth of e-commerce requires robust multimodal representations that capture diverse signals from user-generated listings. Existing vision–language models (VLMs) typically align titles with primary images, i.e., single-view, but overlook non-primary images and auxiliary textual views that provide critical semantics in open marketplaces such as Etsy or Poshmark. To this end, we propose a framework that unifies multimodal and multi-view learning through \textit{Factorized Transport}, a lightweight approximation of optimal transport, designed for scalability and deployment efficiency. During training, the method emphasizes primary views while stochastically sampling auxiliary ones, reducing training cost from quadratic in the number of views to constant per item. At inference, all views are fused into a single cached embedding, preserving the efficiency of two-tower retrieval with no additional online overhead. On an industrial dataset of 1M product listings and 0.3M interactions, our approach delivers consistent improvements in cross-view and query-to-item retrieval, achieving up to +7.9\% Recall@500 over strong multimodal baselines. Overall, our framework bridges scalability with optimal transport–based learning, making multi-view pretraining practical for large-scale e-commerce search.
\end{abstract}




\maketitle

\section{Introduction}
E-commerce continues to grow rapidly, creating new challenges and opportunities in how users interact with, search for, and discover products in online marketplaces~\cite{cheng2010personalized, pancha2022pinnerformer, zhang2024scaling}. Recent advances in vision-language models (VLMs), such as CLIP~\cite{radford2021learning} and ALIGN~\cite{jia2021scaling}, have significantly enhanced cross-modal understanding~\cite{liu2023visual,zhu2025evtp,chen2025prompt}, enabling users to search for products that connect natural language queries with both semantic text and visual product representations. In response, major platforms such as Amazon~\cite{zhu2024bringing}, Walmart~\cite{giahi2025vl}, and DoorDash~\cite{gurjar2025dashclip} have developed proprietary VLMs for semantic search, recommendations, and product understanding.

\begin{figure}
    \centering
    \includegraphics[width=0.82\linewidth]{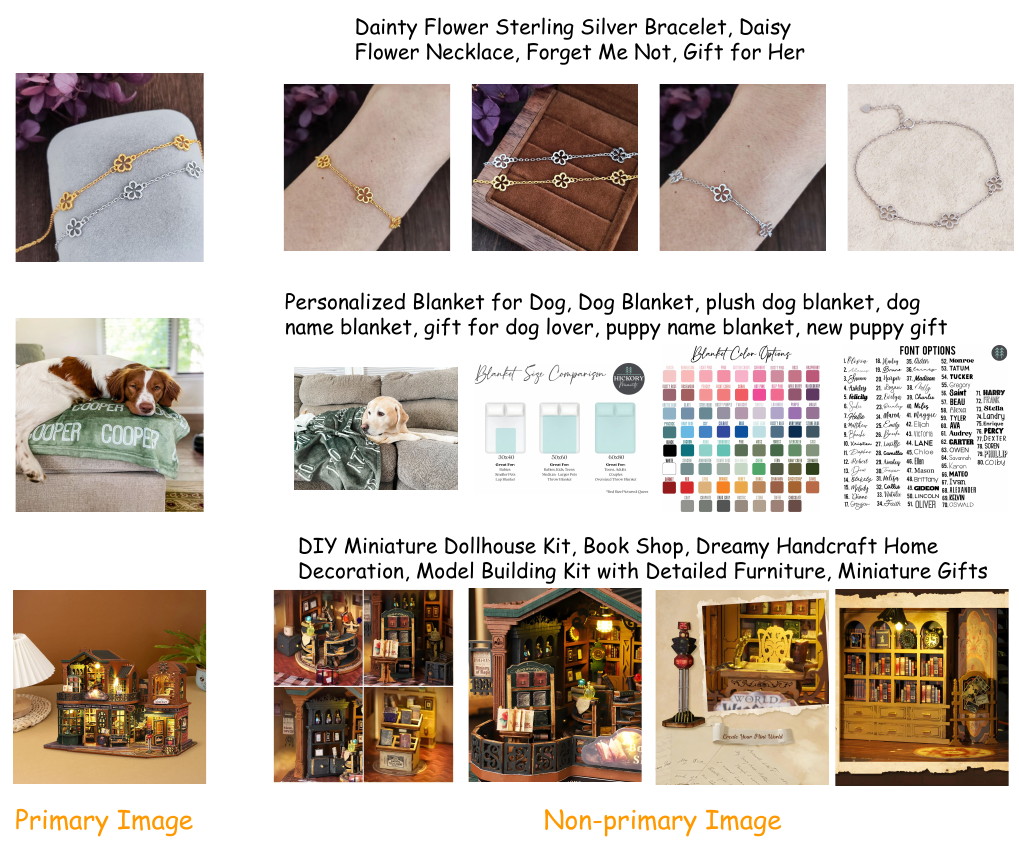}
    \vspace{-0.1in}
    \caption{Example of listings with primary image and non-primary images.}
    \label{fig:example1_1}
\vspace{-0.22in}
\end{figure}

It is worth mentioning that \textit{most existing vision-language pretraining approaches are mainly designed for \textbf{single-view} representation that aligns the product title with the primary image, i.e., the first image of the product}. This setup may be sufficient due to the nature of their platform, since systems benefit from structured metadata, i.e., titles, attributes, and taxonomies, allowing supervised training with well-aligned title–image pairs.
However, this assumption breaks down on platforms like Etsy, Depop, or Poshmark, where listings\footnote{A listing refers to a product item, either physical or digital, that a seller offers for sale.} are heavily customized, highly diverse and lack standardized metadata of both title and primary image, and rich context can come from \textit{\textbf{multi-view}} information. For example, as shown in Fig.~\ref{fig:example1_1}, in the bracelet case, one non-primary image highlights color and material variations, while another shows the bracelet worn, providing context on scale and fit. Similar patterns appear in other listings: the pet blanket includes non-primary images with customizable size and color details, and the dollhouse kit features alternate room configurations and close-ups of decorative elements, which cannot be directly conveyed by the title. These diverse visual cues collectively enrich the semantic representation of the product beyond what the primary image alone can convey. Therefore, ignoring this information not only limits representational coverage but may also reduce downstream retrieval performance in open-ended query scenarios.

\begin{figure*}
    \centering
    \includegraphics[width=0.8\linewidth]{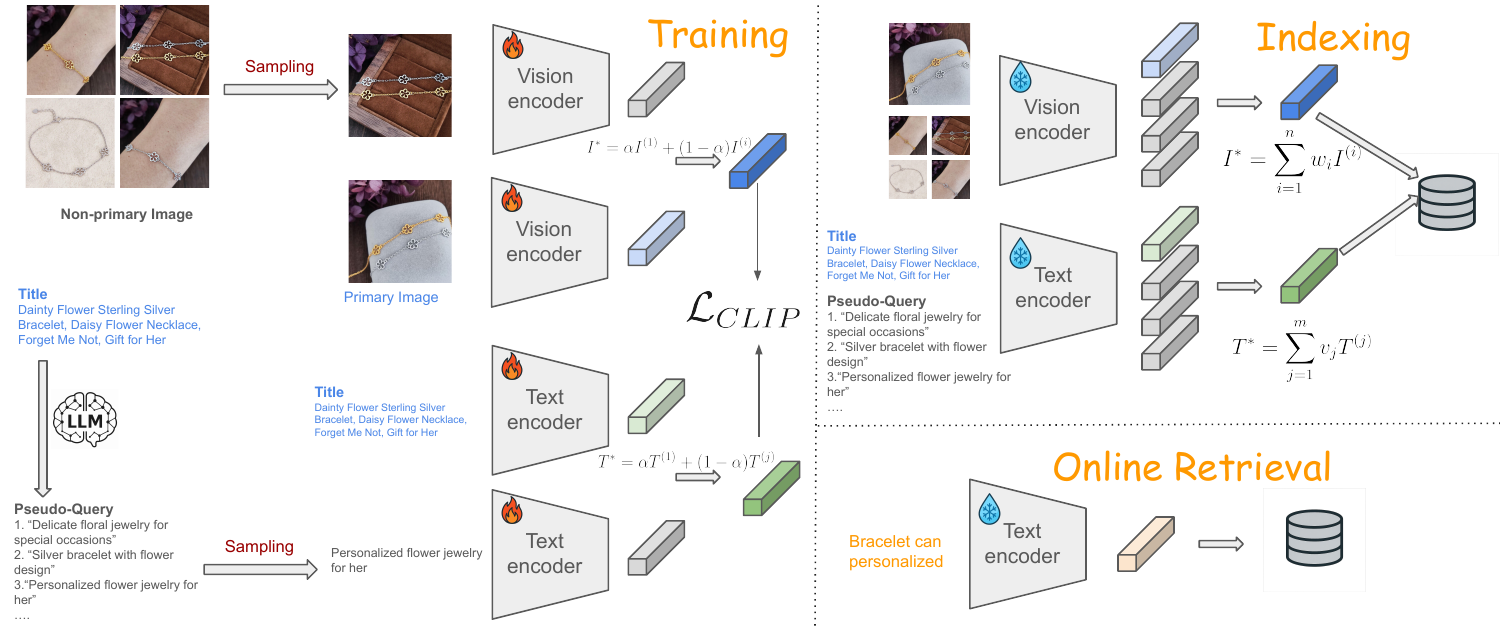}
    \vspace{-0.1in}
    \caption{The proposed framework. During training, primary views (title, primary image) are combined with stochastically sampled auxiliary views (non-primary images, pseudo-queries) to form fused embeddings via a fixed-weight ensemble, where rolling sampling approximates transport with low complexity. At indexing and retrieval time, embeddings from multiple views are aggregated offline into a single representation, enabling efficient online retrieval without additional overhead.}
    \label{fig:framework}
\end{figure*}

 To address these issues, we propose a unified unsupervised framework, grounded in the theory of optimal transport, to improve multimodal and multi-view representation learning for e-commerce. 
This framework involves two phases:  
\begin{itemize}
    \item \textbf{Training phase.} We first generalize CLIP-style contrastive learning by introducing a fixed and factorized transport plan over multi-view image and text inputs. It prioritizes the seller-curated primary views while stochastically sampling auxiliary views during training, constructing fused representations via weighted ensembling. This stochastic approximation retains the essence of factorized transport alignment while reducing complexity from quadratic in the number of image–text views to constant time per listing.
    \item \textbf{Indexing phase.} We apply the same transport-based aggregation to fuse multiple textual or visual views into a single embedding per item. This embedding can be computed offline and cached, introducing no additional overhead during online retrieval or inference.
\end{itemize}
The proposed framework is shown in Fig.~\ref{fig:framework}. Unlike prior works that solve full optimal transport for fine-grained alignment, our approach introduces a lightweight factorized coupling. Methods such as OT-CLIP~\cite{shi2024ot} and AWT~\cite{zhu2024awt} improve vision-language generalization, and Prompt-OT~\cite{chen2025prompt} preserves knowledge via OT regularization, but all suffer from high computational cost. 
By contrast, our method approximates OT with a fixed structure, bridging theoretical alignment benefits with the efficiency needed for scalable real-world e-commerce deployment. Extensive experiments using real-world industrial data on in-domain retrieval and zero-shot downstream tasks demonstrate the effectiveness and scalability of our approach.


\vspace{-0.1in}




\section{Method}

\noindent\textbf{Preliminary.} 
Suppose each listing consists of a set of image views $\mathcal{I} = \{ I^{(i)} \}_{i=1}^n$, and text views $\mathcal{T} = \{ T^{(j)} \}_{j=1}^m$ in \textbf{latent space} obtained by trainable image \( f_{\theta} \) and text encoders \( g_{\phi} \)\, where $I^{(1)}$ and $T^{(1)}$ denote the seller-curated primary image and title, respectively.
For image modality, the remaining views naturally are non-primary images (shown in Fig.~\ref{fig:example1_1}). For text modality, we consider generating pseudo queries, which show advantages in recent literature~\cite{gospodinov2023doc2query,li2024doc2token}. Specifically, we employ \texttt{gpt-4o-mini}~\cite{achiam2023gpt} to generate pseudo-queries based on title information. The prompt template is:

\noindent\makebox[\linewidth][c]{%
\fbox{%
  \parbox{0.95\linewidth}{%
  \textit{You are a creative assistant helping improve search suggestions on an E-commerce Platform. // 
  Given a product title, generate 5--10 short, diverse, and imaginative search queries a curious shopper might use. // 
  Do NOT repeat the title directly. Instead, think about real user needs, use cases, related problems, or benefits. // 
  Each query should be a short phrase --- 3 to 10 words --- like how people search. // 
  Now generate queries for the following product: // 
  Product Title: \{title\} // 
  Search Queries:}
  }%
}%
}
To ensure zero overhead during retrieval, \uline{we aim to learn a \emph{fused} embedding that aggregates multiple views into a single vector per item (per modality).} Diverse views are incorporated during training to enhance alignment, but only one aggregated representation is maintained at inference.
We denote the new embeddings as
\begin{align}\label{eq:0000}
I^* = \sum_{i=1}^{n} w_i I^{(i)}, \qquad
T^* = \sum_{j=1}^{m} v_j T^{(j)}, \qquad
\end{align}
 with some confidence weights $w\!\in\!\Delta^{n}$ and $v\!\in\!\Delta^{m}$. The CLIP loss for a positive pair $(I^*,T^*)$ (omitting the temperature $\tau$ and negatives for clarity) maximizes their inner product:
\begin{align}\label{eq:clip}
\mathcal{L}_{\text{CLIP}} \;\propto\; - \langle I^*, T^* \rangle .
\end{align}\label{eq:l_clip}




\vspace{-0.1in}
\noindent\textbf{Link to Optimal Transport.} Now, we will show how to interpret $\mathcal{L}_{\text{CLIP}} $ in Eq.~\ref{eq:clip} from the perspective of Optimal Transport (OT). We first recap the definition of OT.

\begin{definition}[Optimal Transport Between Multi-View Modalities]
\label{def:vanilla-ot-rephrased}

Let a listing have $n$ image embeddings $\{ I^{(i)} \}_{i=1}^n \subset \mathbb{R}^d$ and $m$ text embeddings $\{ T^{(j)} \}_{j=1}^m \subset \mathbb{R}^d$.  
Define the associated empirical distributions:
\[
\mu_{\mathcal{I}} = \sum_{i=1}^{n} a_i\, \delta_{I^{(i)}}, \qquad
\mu_{\mathcal{T}} = \sum_{j=1}^{m} b_j\, \delta_{T^{(j)}},
\]
where \( a_i, b_j \geq 0 \), \( \sum_i a_i = 1 \), \( \sum_j b_j = 1 \), and \( \delta_x \) denotes a Dirac measure at point \( x \in \mathbb{R}^d \).

The \emph{optimal transport cost} between \( \mu_{\mathcal{I}} \) and \( \mu_{\mathcal{T}} \) under cost function \( c : \mathbb{R}^d \times \mathbb{R}^d \to \mathbb{R}_{\geq 0} \) is:
\[
\mathrm{OT}_c(\mu_{\mathcal{I}}, \mu_{\mathcal{T}}) =
\min_{\gamma \in \Pi(a, b)} \sum_{i=1}^{n} \sum_{j=1}^{m} \gamma_{ij} \cdot c(I^{(i)}, T^{(j)}),
\]
with the transport polytope, i.e., the set of all valid transport couplings $\gamma$:
\[
\Pi(a, b) := \left\{ \gamma \in \mathbb{R}_{\geq 0}^{n \times m} \;\middle|\;
\sum_j \gamma_{ij} = a_i, \quad \sum_i \gamma_{ij} = b_j \right\}.
\]
\end{definition}
\vspace{-0.08in}
We can immediately link $\mathcal{L}_{\text{CLIP}} $ with OT. 
\begin{lemma}[Connection to Factorized Transport Alignment]
\label{lemma:soft-ot}

With a proper choice of modality-specific weights \( w_i \in \Delta^n \) and \( v_j \in \Delta^m \), the contrastive loss in Eq.~\ref{eq:l_clip}, which is based on the dot product of fused embeddings
$
I^* = \sum_{i=1}^n w_i I^{(i)},~ T^* = \sum_{j=1}^m v_j T^{(j)},
$
can be interpreted as a form of optimal transport cost:
\[
\langle I^*, T^* \rangle = \sum_{i=1}^n \sum_{j=1}^m w_i v_j \langle I^{(i)}, T^{(j)} \rangle,
\]
which corresponds to a transport cost under a fixed coupling \( \gamma_{ij} = w_i v_j \) and negative similarity cost \( c(I^{(i)}, T^{(j)}) = -\langle I^{(i)}, T^{(j)} \rangle \).

\end{lemma}



\vspace{-0.1in}
\begin{remark}
A transport coupling that more closely approximates the true optimal transport solution typically leads to better cross-modal alignment.
\end{remark}

\noindent\textbf{Design of Transport Coupling.}
The remark above suggests that a transport coupling closer to the optimal solution can improve modality alignment. In our case, the title and primary image are typically well-aligned and semantically dominant, so they should receive higher weights in the coupling. To reflect this, we assign a fixed weight \( \alpha \in (0, 1) \) to the primary view (e.g., \( w_1 = \alpha \) for images, \( v_1 = \alpha \) for text), and evenly distribute the remaining weight \( 1 - \alpha \) across auxiliary views: \( w_i = \frac{1 - \alpha}{n-1} \) and \( v_j = \frac{1 - \alpha}{m-1} \) for \( i = 2, \ldots, n \), \( j = 2, \ldots, m \). 
This design emphasizes the dominant modalities while remaining simple and generalizable to listings with varying numbers of auxiliary views. It is worth mentioning that this design can be applied for both training and indexing. However, in each step of training, directly using all complementary image and text will cause substantial additional computational overhead. To this end, in the following, we propose to use roll sampling for efficient training.

\noindent\textbf{Training Phase: Rolling Sampling as a Scalable Stochastic Transport Approximation.}
To ensure scalability in large-scale e-commerce pretraining, we adopt a \emph{rolling}, stochastic approximation of the full transport coupling. Computing contrastive alignment over \emph{all} image and text views for every listing has a complexity of \( \mathcal{O}(n \cdot m) \), where \( n \) and \( m \) are the numbers of image and text views, respectively. This becomes impractical at scale, especially when listings contain many auxiliary views such as non-primary images or pseudo-queries. To address this, we employ a \textbf{rolling sampling} strategy: at each training step we randomly draw one primary view (e.g.\ the title or primary image) \emph{and} one auxiliary view (e.g.\ a pseudo-query or non-primary image) from each modality, so that different auxiliary views are visited in successive steps.
Fused embeddings are then constructed via a fixed-weight ensemble
\[
I^* = \alpha I^{(1)} + (1 - \alpha) I^{(i)}, \qquad
T^* = \alpha T^{(1)} + (1 - \alpha) T^{(j)},
\]
where \( i>1 \), \( j>1 \) are randomly sampled from the indice of non-primary images and pseudo-queries, respectively, and \( \alpha\!\in\!(0,1) \) emphasizes primary views.
The dot product
\begin{align*}
    \langle I^*, T^* \rangle
    = &\alpha^2          \langle I^{(1)}, T^{(1)} \rangle
    + \alpha(1-\alpha)  \langle I^{(1)}, T^{(j)} \rangle \\ \nonumber
    + &\alpha(1-\alpha)  \langle I^{(i)}, T^{(1)} \rangle
    + (1-\alpha)^2      \langle I^{(i)}, T^{(j)} \rangle
\end{align*}
is an \emph{unbiased} stochastic estimator of the objective with fixed, factorized coupling \( \gamma_{ij}=w_i v_j \).  

This rolling scheme lowers the per-step complexity from \( \mathcal{O}(n m) \) to \( \mathcal{O}(1) \), enabling efficient multi-view training while preserving distribution-level alignment, which is an essential property for real-world multimodal systems with massive product catalogs.

\noindent\textbf{Indexing Phase.} The indexing phase is simple and efficient. Given the embeddings of both the primary and auxiliary information, we compute the final embedding for each listing using Eq.~\ref{eq:0000}. 
For multi-modal representation, we adopt a simple fusion strategy: for each listing, we take the average of text and image embeddings as the final listing representation. 
This approach avoids introducing any additional cache or storage requirements and incurs no extra computational overhead during future online retrieval.

\vspace{-0.1in}
\section{Experiments}
In this section, we conduct the proof-of-concept analysis on real-world dataset from our e-commerce platform. The dataset is sourced from the database of our e-commerce platform, comprising a random selection of 1,000,000 listings ($m\approx10,n\approx6$ per listing). For user interactions, we sampled a subset of user logs related to the chosen listings in two weeks, which resulted in 292,520 interactions (each involving at least one click) covering ~15\% unique listings. 
The model is initialized from \texttt{Marqo/marqo-ecommerce-embeddings-L}. Training is conducted for 30 epochs using the AdamW optimizer with a learning rate of $1 \times 10^{-5}$. We use a batch size of 128 with gradient accumulation of 4.
Building upon the implementation described above, we next present a series of experiments designed to evaluate the effectiveness and efficiency of our proposed framework. 




 \begin{wraptable}[7]{r}{0.50\linewidth}
\label{tab:my-table}
\resizebox{0.22\textwidth}{!}{%
\begin{tabular}{llc}\toprule
\multicolumn{1}{c}{\textbf{Source}}        & \multicolumn{1}{c}{\textbf{Target}}       & \multicolumn{1}{c}{\textbf{R@1}} \\ \midrule
Non-P Image   & Primary Image & +1.23   \\
Non-P Image   & Title         & +3.69   \\
Primary Image & Non-P Image   & +0.10   \\
Primary Image & Title         & +1.95   \\
Title         & Non-P Image   & +4.36   \\
Title         & Primary Image & +1.90   \\ \bottomrule
\end{tabular}%
}
\end{wraptable}
\vspace{-0.1in}
\subsection{In-domain cross-view Retrieval}
We evaluate retrieval by using one view of an item as the query to retrieve another, covering both intra-modal (image-to-image) and cross-modal (text-to-image) cases. 
As shown in the right table, consistent gains are observed. The largest improvements occur when retrieving between titles and non-primary images (+4.36\% R@1) and between non-primary images and titles (+3.69\% R@1). Smaller but positive effects are also seen for primary-to-title retrieval (+1.95\%) and non-primary to primary image retrieval (+1.23\%). These results confirm that the model effectively captures complementary information across views, improving alignment and overall retrieval quality.

\vspace{-0.1in}
\subsection{Query-to-Item Retrieval}
In real-world e-commerce platforms, two-tower architectures are widely adopted for their scalability. Our approach remains within this paradigm but focuses on enhancing the pre-training phase rather than altering the architecture. We adopt a zero-shot evaluation setup, where the query encoder (text only) is identical to the text encoder of the embedding model, and retrieval is performed with k-NN over fused item embeddings (see Sec .~\ref{sec:multimodal} for analysis). We compare against several embedding baselines, ranging from text-only to multimodal models: \texttt{Qwen-0.6B}~\cite{qwen3embedding}, \texttt{Qwen-8B}~\cite{qwen3embedding}, \texttt{JINAv3}~\cite{sturua2024jina}, \texttt{Gemini-embedding-001}~\cite{lee2025gemini}, \texttt{Marqo-Ecommerce-L}~\cite{zhu2024marqoecommembed_2024}, and \texttt{OpenAI text-embedding-3}~\cite{openai2024textembedding3}. It is noteworthy that \texttt{Marqo} is trained on large-scale multimodal e-commerce data. \texttt{Bertsy} denotes the model finetuned on our in-house data. 

As shown in the table below, our method achieves the best performance across all metrics. At Recall@500, text-only baselines such as \texttt{OpenAI text-embedding-3} and \texttt{Gemini-embedding-001} reach 12.5\% and 13.7\%, respectively, while \texttt{Marqo} achieves 13.1\%. Our single-view multimodal variant improves this to 17.8\%, and incorporating multi-view signals further boosts Recall@500 to 21.0\% (+3.2\% against single-view). Similar improvements are observed at lower cutoffs, with our model reaching +12.7\% at R@10 and +19.1\% at R@100. These results demonstrate that multi-view pre-training substantially enhances query-to-item retrieval, translating into stronger engagement capacity for real e-commerce applications.
\begin{table}[!h]
\vspace{-0.08in}
\vspace{-0.05in}
\label{tab:result-1}
\resizebox{0.35\textwidth}{!}{%
\begin{tabular}{lccc} \toprule
\multicolumn{1}{c}{\textbf{Model}}   & \textbf{R@10} & \textbf{R@100} & \textbf{R@500} \\ \toprule
Qwen-0.6B                       & --        & --         & --        \\
Qwen-8B                         & +4.42        & +9.56         & +11.59        \\
JINAv3                          & +1.77        & +4.72         & +3.86         \\
OpenAI text-embedding-3         & +5.32        & +10.13        & +12.51        \\
Gemini-embedding                & +4.98        & +11.88        & +13.67        \\
Marqo (text-only)               & +5.37        & +7.28         & +6.95         \\
Single-view Bertsy (text-only)  & +8.42        & +12.68        & +14.01        \\
Marqo (multimodal)              & +8.36        & +12.12        & +13.06        \\
Single-view Bertsy (multimodal) & +10.43       & +15.99        & +17.78        \\ \midrule
\textbf{Multi-view Bertsy (multimodal)}  &\textbf{ +12.74}       & \textbf{+19.09 }       & \textbf{+20.97}         \\ \bottomrule
\end{tabular}%
}
\end{table}





\subsection{Training vs Indexing on Taxonomy}
\vspace{-0.05in}
We assess retrieval at the taxonomy level by computing Recall@k per category. As shown in the right figure, 
\begin{wrapfigure}{r}{0.40\linewidth}
  \vspace{-6pt}
  \centering
  \includegraphics[width=\linewidth]{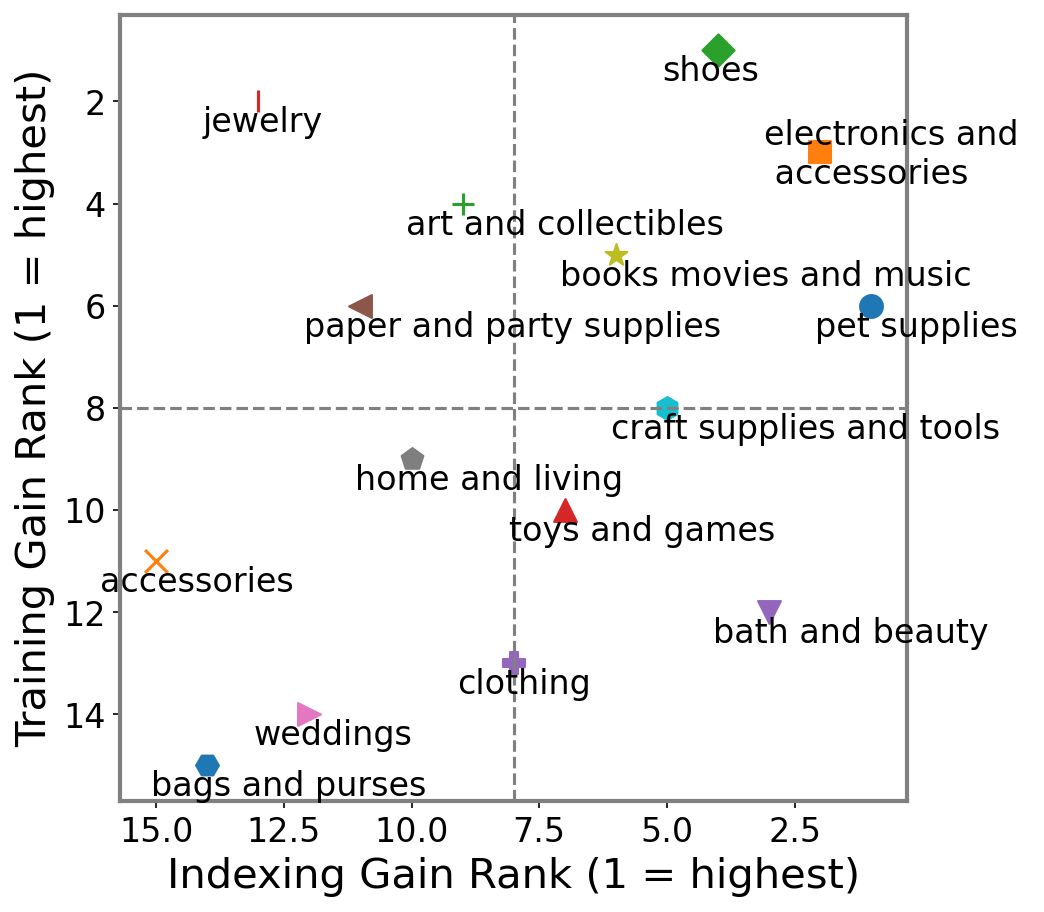}
  \vspace{-0.05in}
\end{wrapfigure}
multi-view training and indexing yield the largest gains for pet supplies, shoes, and electronics, where visual diversity is especially valuable. Jewelry and art benefit mainly from multi-view training, while bath and beauty and toys gain more from multi-view indexing, reflecting the importance of additional perspectives at retrieval time. In contrast, categories such as bags, accessories, and weddings show minimal improvement, consistent with their reliance on a single representative image. This result suggests taxonomy may be a strong prior for representation learning, worth exploring in future work.

\begin{wraptable}[4]{r}{0.4\linewidth}
\label{tab:111}
\resizebox{0.2\textwidth}{!}{%
\begin{tabular}{ccccc}\toprule
\textbf{Text} & \textbf{Image} & \textbf{R@10} & \textbf{R@100} & \textbf{R@500} \\ \midrule
\ding{51}             & \ding{55}              & -1.65       & -2.03        & -2.12        \\
\ding{55}             & \ding{51}              & -8.84       & -11.32        & -11.23 \\ \bottomrule      
\end{tabular}%
}
\end{wraptable}
\vspace{-0.09in}
\subsection{Multimodal Representation}\label{sec:multimodal}
\vspace{-0.04in}
Recap that our framework applies average pooling, that is, element-wise averaging, to fuse the two modalities into a single vector for each item. To validate the effectiveness of multimodal representations, we evaluate the contribution of both text and image embeddings by removing them individually. The results in the right table show that excluding the text modality leads to a performance drop of approximately 2\%, while excluding the image modality results in a more substantial decline of about 11\%. These findings confirm that incorporating both modalities is beneficial, and leveraging multimodal representations yields more robust and informative item embeddings.

\vspace{-0.09in}
\subsection{Efficiency}
\vspace{-0.05in}
\begin{wraptable}[4]{r}{0.4\linewidth}
\label{tab:efficiency}
\resizebox{0.2\textwidth}{!}{%
\begin{tabular}{ccc} \toprule
\textbf{seconds}     & \textbf{GPU} & \textbf{Training Time} \\ \midrule
\textbf{Single View} & 25GB/device  & 19 s/step             \\
\textbf{Multi View}  & 29GB/device  & 31 s/step     \\ \bottomrule        
\end{tabular}%
}
\end{wraptable}
Although incorporating auxiliary views expands the training data to over six times the size of the original single-view setup, our method achieves efficiency by employing a rolling sampling strategy to approximate the optimal transport alignment. As shown in the right table, this results in only modest computational overhead, with GPU memory increasing from 16\% per device and training time from 19s to 31s per step. Thus, our framework effectively fuses multi-view and multimodal signals while maintaining scalability and efficiency.

\vspace{-0.1in}
\section{Conclusion}
\vspace{-0.03in}
We introduced a factorized transport alignment framework that bridges theoretically principled optimal transport with the practical constraints of industrial-scale e-commerce, achieving both strong alignment and deployment efficiency.
Our method efficiently fuses primary and auxiliary views while remaining compatible with CLIP-style training. Experiments on industrial-scale data show consistent gains in cross-view and zero-shot query-to-item retrieval against existing single-view multimodal models. Beyond performance, the approach incurs only modest overhead, making it practical for real-world deployment. Future work may extend this framework to personalization and reinforcement from user feedback.

\newpage
\section*{Ethical Considerations}
Our work focuses on improving multimodal and multi-view representation learning for e-commerce retrieval. While the framework is designed to enhance search quality and scalability, several ethical considerations arise. First, user-generated data on open marketplaces may contain sensitive or personally identifiable information; therefore, appropriate anonymization and filtering are critical before training. Second, as with many embedding-based models, there is a risk of misuse if the technology is applied beyond intended domains, such as surveillance or profiling. To mitigate these concerns, we rely on large-scale, anonymized product listings rather than personal data, and we highlight the importance of responsible deployment practices. Overall, the framework is intended to advance retrieval quality while remaining mindful of privacy, security, and potential misuse.

\bibliographystyle{ACM-Reference-Format}
\balance
\bibliography{myref.bib}


\end{document}